\documentclass[reprint,pra]{revtex4-1}

\usepackage{amsmath}
\usepackage{amssymb}
\usepackage{graphicx}
\usepackage{dcolumn}
\usepackage{bm}
\usepackage{braket}
\usepackage{url}
\usepackage[breaklinks]{hyperref}
\hypersetup{colorlinks,citecolor=cyan,urlcolor=blue}

\begin{document}

\title{Numerical linked cluster expansions for inhomogeneous systems}

\author{Johann Gan}
\email{johanngan.us@gmail.com}

\author{Kaden R. A. Hazzard}
\email{kaden.hazzard@gmail.com}
\affiliation{Department of Physics and Astronomy, Rice University, Houston, Texas 77005, USA}

\date{May 6, 2020}

\begin{abstract}
	We develop a numerical linked cluster expansion (NLCE) method that can be applied directly to inhomogeneous systems, for example Hamiltonians with disorder and dynamics initiated from inhomogeneous initial states. We demonstrate the method by calculating dynamics for single-spin expectations and spin correlations in two-dimensional spin models on a square lattice, starting from a checkerboard state. We show that NLCE can give moderate to dramatic improvement over an exact diagonalization of comparable computational cost, and that the advantage in computational resources grows exponentially as the size of the clusters included grows. Although the method applies to any type of NLCE, our explicit benchmarks use the rectangle expansion. Besides showing the capability to treat inhomogeneous systems, these benchmarks demonstrate the rectangle expansion's utility out of equilibrium.
\end{abstract}

\maketitle

\section{\label{sec:intro}Introduction}

	The presence of inhomogeneity in a quantum system can drastically alter its properties. One well-known example is Anderson localization, where a particle in a disordered lattice fails to diffuse, instead remaining trapped in a localized region of space~\cite{anderson1958}. More recently, there has been significant interest in a counterpart to Anderson localization in interacting many-body systems~\cite{parameswaran2018,nandkishore2015,abanin2019,alet2018}. Contrary to traditional assumptions of statistical physics, such many-body localized systems can preserve spatial inhomogeneities from their initial condition, never fully reaching thermal equilibrium. Studying these systems could lead to fundamental advancements in statistical physics, exotic materials like Floquet time crystals~\cite{choi2017,zhang2017,wilczek2012}, and stabilizing mechanisms for quantum memory~\cite{goihl2020,goihl2019,bahri2015}.

	Unfortunately, theory is lagging behind experiment for disordered quantum systems. Due to superposition, the Hilbert space dimension of a quantum system increases exponentially with system size. The lack of symmetry in inhomogeneous systems makes computations even more challenging. Various state-of-the-art numerical methods exist, but they are usually limited in applicability. For example, tensor network methods have had widespread success in one and two dimensions~\cite{orus2019}, including in one-dimensional (1D) disordered systems~\cite{chandran2015,pollmann2016,wahl2017,goldsborough2017}, but the area law of entanglement and the apparent need for large bond dimensions in three dimensions has thus far hindered application in higher dimensions~\cite{hauru2018}. Another example is quantum Monte Carlo (QMC) simulation~\cite{li2019,zhang2019,pollet2013,pollet2012,gull2011,scalettar1999}. QMC is not as limited by dimension but suffers from the sign problem in many systems, including systems with fermions, frustrated systems, and for dynamics. A third example is the nonequilibrium Green's function approach, which has been shown to work well for out-of-equilibrium systems in any number of spatial dimensions, but only in the regime of weak to moderate interactions~\cite{schlunzen2017,negf2016}. There are also a number of other methods, each with unique limitations. Exact diagonalization (ED) remains one of the only truly general tools for many-body quantum simulation, but it also rapidly becomes infeasible as system size grows. Improved numerics is needed for studies of inhomogeneous systems, especially in higher dimensions.

	One technique that has seen much use in translationally invariant systems is the numerical linked cluster expansion (NLCE). NLCE has been widely used to calculate equilibrium properties in uniform, infinite lattices, for many common models such as spin models~\cite{rigol2006,rigol2007a,kallin2013,stoudenmire2014,sherman2016} and fermionic models~\cite{rigol2007b,khatami201111,cheuk2016,khatami2016,brown2017}. It is the method of choice for treating some systems of immense importance, for example the strong-coupling, finite-temperature Fermi-Hubbard model, which is ubiquitous in ultracold matter~\cite{hart2015,brown2017,nichols2018}. Recently, NLCE has also been demonstrated in a wide range of nonequilibrium scenarios, including dynamics~\cite{white2017,mallayya2018,guardadosanchez2018,mallayya201905}, long-time results after a quench~\cite{rigol2014a,rigol2014b,wouters2014,piroli2017,mallayya2017}, driven-unitary systems~\cite{mallayya201907}, and driven-dissipative systems at steady-state~\cite{biella2018}.

	Although NLCE has been extended to disordered systems, it has faced serious difficulties. Refs.~\cite{tang201504,tang201505} showed that NLCE can be used in systems with binary disorder, and Ref.~\cite{mulanix2019} extended this to approximating continuous disorder via carefully chosen discrete disorder levels. However, these methods are computationally expensive, as they require averaging over all possible disorder configurations and thus incur a cost that grows exponentially with system size. This cost multiplies the already exponentially growing cost of NLCE in uniform systems.

	In this paper, we introduce a generalized NLCE algorithm that can be used to calculate local properties on arbitrary inhomogeneous lattices, which allows one to treat discrete or continuous disorder without exponentially expensive averaging, and also allows for spatial inhomogeneities in the initial state for dynamics. This approach has been used in a limited capacity by Devakul and Singh in Ref.~\cite{devakul2015} to compute ground-state entanglement entropy in disordered systems, but the full extent of its applicability to local observables and to dynamics with inhomogeneous initial conditions has, to our knowledge, never been recognized. We focus on dynamics, but the ideas apply to equilibrium and steady-state calculations as well. Incidentally, our primary NLCE uses a rectangle-based expansion~\cite{kallin2013}, which offers a favorable tradeoff between accuracy and efficiency, and the calculations in this paper demonstrate its applicability to dynamics.

	Section~\ref{sec:methods} discusses the numerical methods used in this paper, including the NLCE algorithm and its variants, as well as the models we use for benchmarking. In particular, Sec.~\ref{sec:inlce} introduces the main result of this paper: the inhomogeneous NLCE. Section~\ref{sec:results} presents the results of our benchmark tests for the inhomogeneous NLCE, primarily comparisons to ED for a variety of dynamics calculations in inhomogeneous conditions. Section~\ref{sec:conclusions} concludes and offers potential routes of improvement for the inhomogeneous NLCE.

\section{\label{sec:methods}Methods}
\subsection{\label{sec:hnlce}Homogeneous NLCE}
	NLCEs are a class of methods introduced in Refs.~\cite{rigol2006,rigol2007a,rigol2007b} for approximating observables on an infinite lattice by combining results from a collection of finite subclusters. This method exactly reproduces high-temperature series results (or short-time expansion results, in the case of dynamics) to an order determined by the class of clusters that are included, but has better convergence properties because the clusters are solved exactly rather than perturbatively. Ref.~\cite{tang2013} provides a pedagogical introduction.
	
	Traditionally, NLCE has been used to compute the average value of an extensive observable $A$ in a thermodynamically large lattice $\mathcal L$. For a subcluster $c \subseteq \mathcal L$, define 
	\begin{equation}
		\label{eq:homogeneousproperty}
		P(c) \equiv \sum_{i \in c} \Braket{A_i}_c,
	\end{equation}
	where the local observable $\Braket{A_i}_c$ is summed over all sites $i$ in $c$, and the expectation value is taken in the isolated cluster $c$, i.e. assuming no interactions between sites in $c$ and outside of $c$. The observable of interest is the average value over all $N$ sites in the lattice,
	\begin{equation}
		\label{eq:latticeavg}
		\overline A \equiv P(\mathcal L)/N.
	\end{equation}

	The key idea of NLCE is to treat $\overline A$ as a sum of contributions from all possible subclusters. To do this, we first define the contribution or weight from cluster $c$, $W(c)$, recursively by
	\begin{equation}
		\label{eq:homogeneousweight}
		W(c) \equiv P(c) - \sum_{s \subset c}W(s),
	\end{equation}
	with the base case being a single-site cluster $c_1$ where $W(c_1) = P(c_1)$.
	
	Intuitively, $W(c)$ represents the ``nonadditive'' correction to the true property value $P(c)$ not accounted for by simply summing effects from smaller subclusters within $c$. Hence, if $c$ has $n$ sites, $W(c)$ can be interpreted as a contribution to $P(c)$ that arises from $n$-body correlations in $c$. This has two important consequences. Firstly, $W(c)$ can be nonzero only if $c$ is a connected subcluster, since there cannot be $n$-body correlations if some sites are isolated from others. This can be proven rigorously by induction for a cluster $c$ assembled from two disjoint subclusters $c_1$ and $c_2$, using the fact that
	\begin{equation}
		\label{eq:unconnected}
		P(c) = P(c_1) + P(c_2).
	\end{equation}
	The proof is sketched in Ref.~\cite{tang2013}. Thus, we only need to include linked clusters in the sum in Eq.~\eqref{eq:homogeneousweight}. Secondly, for a system with correlations that decay rapidly with separation outside of some correlation length, we can expect that $W(c) \rightarrow 0$ quickly as $c$ becomes large.

	Rearranging terms in Eq.~\eqref{eq:homogeneousweight} gives
	\begin{equation}
		\label{eq:homogeneousweightsum}
		P(c) = \sum_{s \subseteq c}W(s).
	\end{equation}
	In the limit $c \rightarrow \mathcal L$, we recover an exact infinite summation for $P(\mathcal L)$. This leads to the NLCE, which approximates this infinite series with a finite truncation by some metric of cluster size $|c|$ (to be described momentarily),
	\begin{equation}
		\label{eq:homogeneousnlce}
		\overline A = \frac{1}{N}P(\mathcal L) \approx \sum_{|c| \leq n}W(c),
	\end{equation}
	where each cluster shape $c$ is included in this sum only once---not each of its translations. Since each shape $c$ has $N$ possible translations (up to subleading corrections for $N\rightarrow \infty$), this leads to the $1/N$ multiplying $P(\mathcal L)$. The sum in Eq.~\eqref{eq:homogeneousnlce} usually converges rapidly because typically $W(c)\rightarrow 0$ rapidly as $|c|\rightarrow \infty$. We note that for efficiency, one may use symmetry or topological equivalence to save from solving equivalent clusters multiple times.

	This truncation remains reasonable so long as the sum of weights from truncated clusters is small. Heuristically, an NLCE of order $n$ (which involves $\sim n$ sites) can capture correlations between sites separated by a distance $\mathcal O(n)$, whereas ED requires $n^d$ sites where $d$ is the number of spatial dimensions. In equilibrium calculations, the order required usually grows as temperature is lowered, since increasingly long-range correlations can exist~\cite{tang2013,rigol2006}. In dynamics calculations, the same is true as time grows, and correlations have been able to spread across the system~\cite{white2017}.

\subsection{\label{sec:rectangle}Rectangle expansion}
	Equation~\eqref{eq:homogeneousnlce} can give rise to various summations, depending on how one defines the class of clusters to sum over, which is mostly arbitrary as long as it is done consistently in Eqs.~\eqref{eq:homogeneousweight} and \eqref{eq:homogeneousnlce}. One approach is a site-based expansion, which allows all connected subclusters of the lattice. For models with next-nearest-neighbor (or even longer-ranged) interactions, a similar approach involves restricting subclusters to be maximally connected~\cite{mallayya2017}. A simpler approach on square lattices is the rectangle expansion~\cite{kallin2013}, which allows only rectangular clusters of the shape $s_1 \times \dots \times s_d$ in $d$ dimensions. There are also approaches based on other enumerations like plaquettes~\cite{khatami201104,khatami201112,khatami2012,biella2018,mulanix2019}, which allow geometries built from a unit cell larger than one site, such as a $2\times 2$ square.

	The site expansion is a natural choice of NLCE, and a site expansion including terms up to $n$ sites is likely the most accurate enumeration among those that include clusters of $n$ sites or less, since it includes the most information about subclusters. However, the site expansion has drawbacks due to the exponential~\cite{jensen2000} growth of the number of clusters with the number of sites involved. At higher orders, enumerating the clusters becomes prohibitively expensive, and the computational cost of simulating so many subclusters starts to outweigh the benefits over a simple ED. Furthermore, the large number of clusters leads to summations with many terms, which can cause numerically instability. In practice, the number of clusters can be reduced due to topological equivalences between them~\cite{tang2013}, but this only delays the problem.

	Although our general method for calculating properties in inhomogeneous systems applies to any NLCE enumeration scheme, our numerical results in this paper focus on the rectangle expansion, which circumvents the aforementioned issues with the site expansion at the expense of slower order-by-order convergence. We quantify rectangle size by average side length and define an NLCE of order $n$ to include contributions from rectangles of sizes no greater than $n$.

\subsection{\label{sec:inlce}Inhomogeneous NLCE}
	We now describe a natural extension to NLCE in systems that break translational invariance, either through the system Hamiltonian or through the initial conditions in a dynamics calculation.

	Consider an arbitrary lattice where every site is potentially unique, in which we are interested in measuring a local observable $\Braket{A_M}$ supported on a finite set of sites $M$. For example, for a spin-$1/2$ system, if $A_M$ is $\sigma_k^z$, the usual $z$-component of spin at site $k$, then $M = \{k\}$. If $A_M$ is a two-point correlation between sites $i$ and $j$, then $M = \{i, j\}$.

	For a cluster shape $c$, define an extensive (in the cluster size $|c|$, for large $|c|$) property
	\begin{equation}
		\label{eq:inhomogeneousproperty}
		P(c) = \sum_{t \in T_M(c)} \Braket{A_M}_t,
	\end{equation}
	where $T_M(c)$ is the set of translations $t$ of $c$ within $\mathcal L$ such that $M \subseteq t$, and $\Braket{A_M}_t$ is the expectation of $A_M$ as computed on the isolated cluster $t$. A key consequence of this definition is that $P(\mathcal L)/N = \Braket{A_M}$, with the expectation value taken on the full lattice $\mathcal L$; the quantity on the right-hand side is what we wish to calculate.

	\begin{figure}[htb]
		\includegraphics[width=.48\textwidth,trim={.3cm 3.5cm 1.3cm 1.5cm},clip]{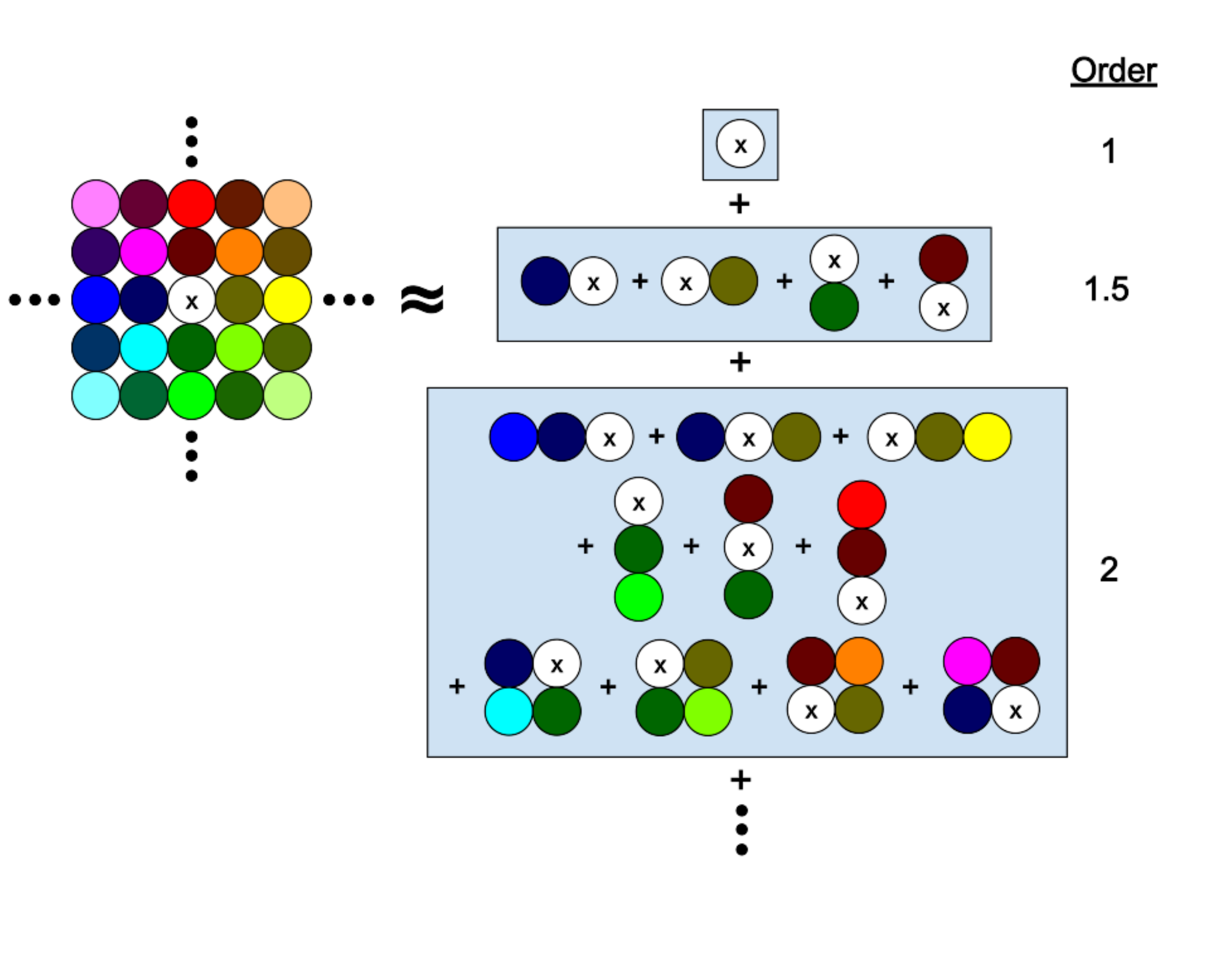}
		\caption{\label{fig:irnlce} NLCE for a single-site measurement in an inhomogeneous system, using the rectangle expansion. For each cluster shape, weights are computed for every possible translation that contains the measured site. Weights are summed as in the homogeneous NLCE.}
	\end{figure}

	With this definition of $P(c)$, we can define $W(c)$ as in Eq.~\eqref{eq:homogeneousweight}. Note that Eq.~\eqref{eq:unconnected} remains valid under the definition of $P(c)$ in Eq.~\eqref{eq:inhomogeneousproperty}, meaning that the weights of unconnected clusters vanish in the inhomogeneous case, just as in the homogeneous case (the proof is the same). Thus the NLCE approximation becomes
	\begin{equation}
		\label{eq:inhomogeneousnlce}
		\Braket{A_M} \approx \sum_{|c| \leq n}W(c).
	\end{equation}

	Figure~\ref{fig:irnlce} depicts the inhomogeneous NLCE, using the rectangle expansion. The right-hand side of Eq.~\eqref{eq:inhomogeneousnlce} is the same as in Eq.~\eqref{eq:homogeneousnlce}, but the quantity being approximated can be an arbitrary observable instead of just the average value of an extensive observable. For a homogeneous lattice, this formula reduces to the homogeneous NLCE. We focus on dynamics in this paper, but Eq.~\eqref{eq:inhomogeneousnlce} can just as easily be used for equilibrium and steady-state calculations.

	The flexibility to handle inequivalent sites in inhomogeneous systems comes at the price of needing to simulate more clusters per NLCE order, since translated clusters of the same shape are no longer identical. Furthermore, topological and symmetry equivalences can no longer be exploited to reduce the number of clusters in general (although if the inhomogeneous system still has some symmetries, they can still be utilized). For the rectangle expansion, if the largest cluster contains $N$ sites, then the inhomogeneous NLCE is roughly $N$ times as expensive as the homogeneous NLCE. For the site expansion, this factor is greater than $N$ due to the loss of topological equivalence.
	
	Despite the extra cost of the inhomogeneous NLCE, it still provides an exponential improvement over previous NLCE methods for simulating disorder~\cite{tang201504,tang201505,mulanix2019}, which require averaging over all disorder realizations and are $m^N$ times as expensive as the homogeneous NLCE given $m$ discrete disorder levels. Additionally, the cost is somewhat mitigated by the NLCE being trivially parallelizable, regardless of the expansion used.

	Often the \textit{average} under many disorder realizations is the quantity of interest~\cite{alet2018}. Since one can compute single-realization behavior with the inhomogeneous NLCE, it is straightforward to calculate both disorder averages and statistical error estimates. Achieving a relative statistical error $\epsilon$ in the disorder average requires only a finite number, $\mathcal O(1/\epsilon^2)$, of realizations, so does not impact the exponential improvement achieved by this method.

	\subsection{\label{sec:disorder}Models}
\begin{figure}[htb]
		\includegraphics[width=.48\textwidth,trim={3.4cm .2cm 2.3cm 0},clip]{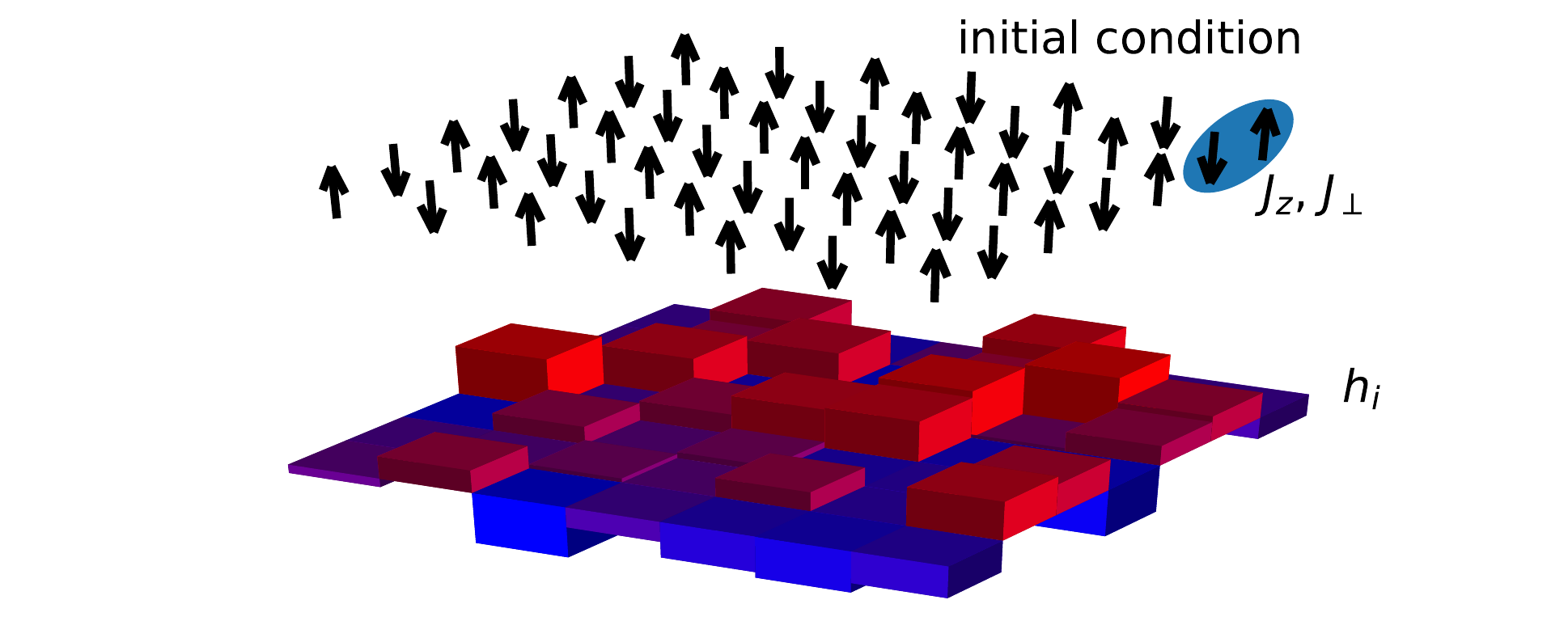}
		\caption{\label{fig:modelcartoon} Illustration of one physical scenario we use to test and benchmark the inhomogeneous NLCE. A spin-1/2 square lattice is initialized in a checkerboard state, and then time evolved with XXZ interactions and a disordered longitudinal field, $h_i$, applied at each site $i$. The $h_i$ are drawn from a continuous uniform distribution on the range $[-D, +D]$, for some disorder strength $D$.}
	\end{figure}

	We benchmark and illustrate the inhomogeneous NLCE by computing time dynamics in two models. We have chosen these models primarily for their simplicity and physical relevance, and they also lead to interesting behaviors in their own right. We focus on two-dimensional (2D) square lattices for both models, with brief comparisons to 1D chains. The scenario is depicted in Fig.~\ref{fig:modelcartoon}.

	The first model is the nearest-neighbor XXZ model with continuous disorder, a prototypical model for studies of many-body localization,
	\begin{equation}
		\label{eq:disorderedxxz}
		\hat H_\text{XXZ} = -\sum_{\Braket{i,j}}\left[J_\perp\left(\sigma_i^x\sigma_j^x + \sigma_i^y\sigma_j^y\right) + J_z\sigma_i^z\sigma_j^z\right] + \sum_i h_i\sigma_i^z,
	\end{equation}
	where the disordered field values $h_i$ are drawn from a continuous uniform distribution between $[-D, D]$. Various values of disorder strength $D$ are investigated. For $J_z = 0$, this model can be mapped to the hardcore (strongly interacting) limit of a disordered Bose-Hubbard model, which has been realized in ultracold atom experiments~\cite{fallani2007,meldgin2016,choi2016,lukin2019}. In 1D, it can also be mapped to a system of interacting spinless fermions via a Jordan-Wigner transformation~\cite{alet2018}. This disordered XXZ model is believed to have a many-body localized phase at strong enough disorder~\cite{znidaric2008,canovi2011,deluca2013,serbyn201306,serbyn201309}.

	\begin{figure*}[ht]
		\includegraphics[width=\textwidth]{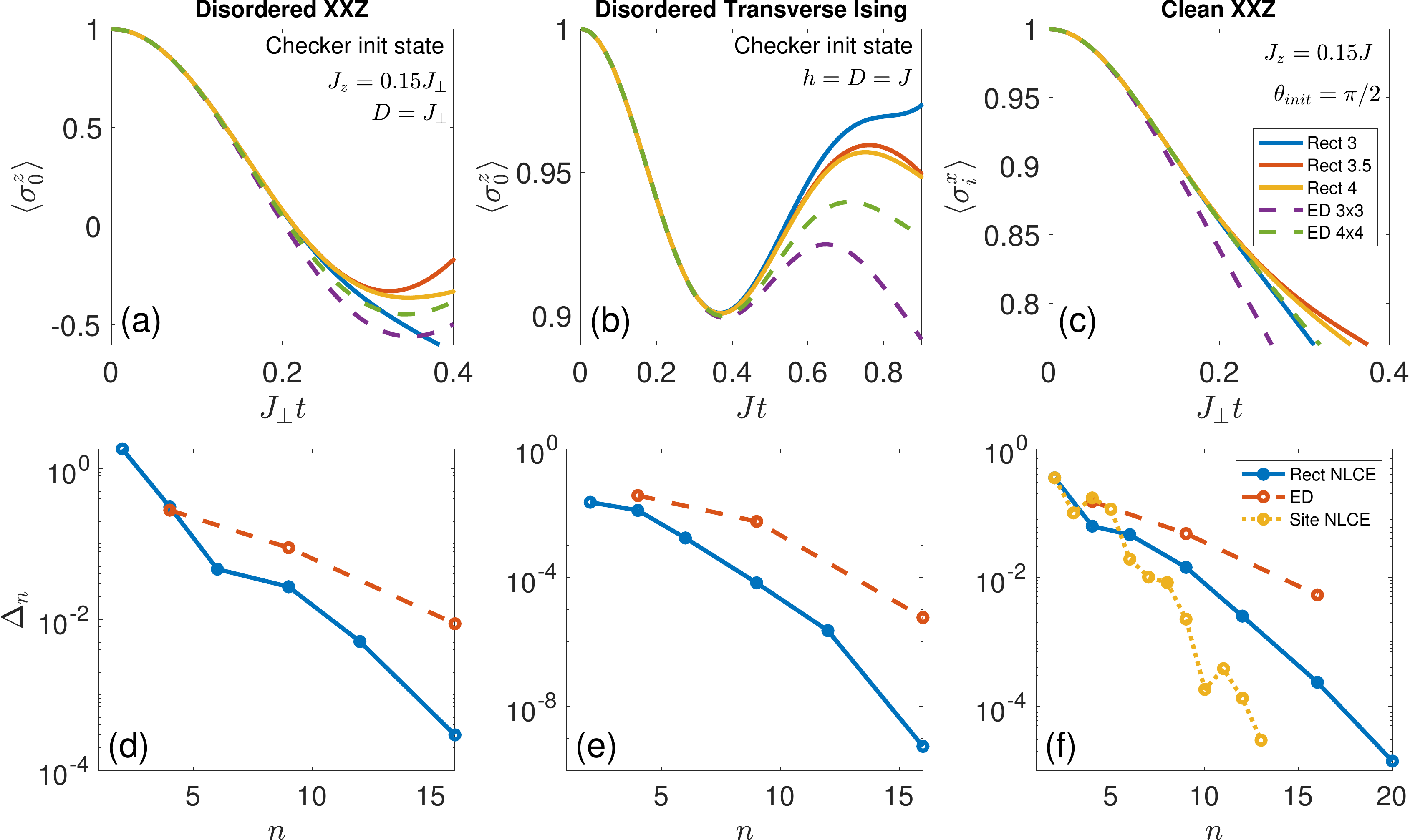}
		\caption{\label{fig:sigma} (a)-(c): Time evolution of a single spin in various models, computed by rectangle NLCE of orders 3, 3.5, and 4 [solid blue, orange, and yellow: bottom-top-middle in (a) and (c), top-to-bottom in (b)]. EDs from $3\times 3$ and $4\times 4$ lattices (dashed purple and green: bottom-to-top) are provided for reference. (a) $\hat H_\text{XXZ}$ with $J_z = 0.15J_\perp$ and $D = J_\perp$, measuring $\Braket{\sigma_0^z}$ where site $0$ is associated with $\Ket{+1}$ in the checkerboard pattern. (b) $\hat H_\text{Ising}$ with $h = D = J$, with the same measurement as (a). (c) $\hat H_\text{XXZ}$ with $J_z = 0.15J_\perp$ and $D = 0$, with initial state $\Ket{\psi(0)} = \bigotimes_i \Ket{+\bm x}_i$, measuring $\Braket{\sigma_i^x}$ for one site. (d)-(f): Proxy for finite size error $\Delta_n$ in (a)-(c) for rectangle NLCE (solid blue), ED (dashed orange), and site NLCE (dotted yellow). Since each NLCE uses multiple EDs of various sizes, $n$ is defined as the number of sites in the largest ED used. $\Delta_n$ is the absolute difference between approximations at order $n$ and the previous order at fixed time $J_\perp t, Jt = 0.25$, normalized by the change in $n$.}
	\end{figure*}

	The second model is the nearest-neighbor transverse Ising model with continuous longitudinal disorder,
	\begin{equation}
		\label{eq:disorderedtransverseising}
		\hat H_\text{Ising} = -J\sum_{\Braket{i,j}}\sigma_i^z\sigma_j^z - h\sum_i\sigma_i^x + \sum_ih_i\sigma_i^z.
	\end{equation}
	Ising models are well-studied in condensed matter physics since they have a relatively simple form but can display a wide variety of interesting phenomena. For example, the model above, which has been realized experimentally in the insulating, dipole-coupled Ising magnet LiHo$_x$Y$_{1-x}$F$_4$~\cite{jonsson2007,quilliam2012,silevitch2019}, is generally thought to exhibit a quantum spin glass phase, although this has recently been the subject of much debate~\cite{gingras2011,mydosh2015}.

	We consider spin dynamics from some initial quantum state when evolved under one of the aforementioned Hamiltonians. Unless explicitly noted, the initial state is a spatially nonuniform, checkerboard product state
	\begin{equation}
		\label{eq:checkerboardstate}
		\Ket{\psi(0)} = \bigotimes_i\Ket{(-1)^{x_i+y_i}}_i,
	\end{equation}
	where $\sigma_i^z\Ket{\pm 1}_i = \pm 1\Ket{\pm 1}_i$. Note that since this state breaks translational symmetry, it is incompatible with the traditional NLCE. Furthermore, there is nothing special about this state from the point of view of the inhomogeneous NLCE; any inhomogeneous initial condition can be accomodated, for example domain walls or random configurations. We will also briefly show results initiated from uniform product states for comparison.

\section{\label{sec:results}Benchmark results}

	Figures~\ref{fig:sigma}(a)-(b) show the main numerical results of this work, demonstrating the validity of the NLCE derived in Sec.~\ref{sec:inlce} and showing its increased accuracy relative to ED calculations requiring solution of clusters up to a similar number of sites. Figure~\ref{fig:sigma}(a) shows time dynamics for $\Braket{\sigma^z}$ at a single site when evolved under a single realization of $\hat H_\text{XXZ}$, starting from the checkerboard initial state. Note that an NLCE of order 4 is comparable in computational cost to a $4\times 4$ ED, since the largest diagonalized system in each contains 16 sites. The times at which these two curves diverge from the corresponding next-highest order results (order $3.5$ NLCE and $3\times 3$ ED, respectively) indicate the accuracy of the two methods. The 4th order NLCE closely agrees with order 3.5 for times until $J_\perp t \sim 0.25$, suggesting that it is accurate for (at least) $J_\perp t \lesssim 0.25$. On the other hand, the $4\times 4$ ED diverges from the $3\times 3$ ED at $J_\perp t \sim 0.15$, suggesting that it is less accurate than the 4th order NLCE. In fact, the first deviations occur later for the order 3 NLCE than for the $4\times 4$ ED, even though this NLCE requires solving at most a 9-site cluster, which requires significantly fewer computational resources than the 16-site ED (roughly a factor of $2^7 \sim 100$ in time and memory). Figure~\ref{fig:sigma}(b) show similar results, but for $\hat H_\text{Ising}$. We stress that although NLCE involves solving multiple clusters instead of just one cluster as in ED, the incurred cost is relatively modest because it is trivially parallelizable and also rapidly compensated for by the reduced $n$ needed by NLCE, which provides an exponential reduction in computational cost.
	
	Figure~\ref{fig:sigma}(c) demonstrates a conclusion tangential to the main focus of this paper: in the traditional \textit{homogeneous} case, the rectangle expansion NLCE is still applicable to dynamics calculations, and can be more accurate than ED. Although the rectangle expansion has been demonstrated in equilibrium~\cite{kallin2013}, NLCEs for dynamics have thus far relied solely on site expansions and maximally connected expansions~\cite{white2017,mallayya2018,guardadosanchez2018}. The panel shows time dynamics for $\Braket{\sigma_i^x}$ in a uniform system initialized in the state $\Ket{\psi(0)} = \bigotimes_i \Ket{+\bm x}_i$ (where $\sigma_i^x\Ket{+\bm x}_i = \Ket{+\bm x}_i$) and evolved under $\hat H_\text{XXZ}$ with $D = 0$. Like in Figs.~\ref{fig:sigma}(a)-(b), NLCE visibly outperforms ED.

	Figures~\ref{fig:sigma}(d)-(f) quantify the convergence of results in Figs.~\ref{fig:sigma}(a)-(c). As each NLCE contains clusters of varying sizes, we define $n$ as the maximum number of sites over all clusters. $\Delta_n$ is a proxy for finite size error, defined as
	\begin{equation}
		\label{eq:delta}
		\Delta_n = \frac{\Braket{\sigma(\tau = 0.25)}_n - \Braket{\sigma(\tau = 0.25)}_{n_\text{prev}}}{n - n_\text{prev}},
	\end{equation}
	where $\sigma$ is $\sigma_0^z$ or $\sigma_i^x$, $\tau$ is $J_\perp t$ or $Jt$, $n_\text{prev}$ is the value of $n$ at the previous order, and $\Braket{\sigma}_n$ is the estimate for $\Braket{\sigma}$ generated by the size-$n$ approximation. We also tried other methods for estimating error (including fitting $\Braket{\sigma}_n$ to a true value plus an exponentially decaying error term, and different normalizations of $\Delta_n$ with the intervals between $n$), but the conclusions are robust. In all three systems, the NLCE converges faster than ED. Figure~\ref{fig:sigma}(f) confirms that the rectangle expansion is in between the site expansion and ED in performance. These conclusions are not special to $\tau = 0.25$, and hold at other times up to those at which the methods become inaccurate.

	\begin{figure}[b]
		\includegraphics[width=.48\textwidth]{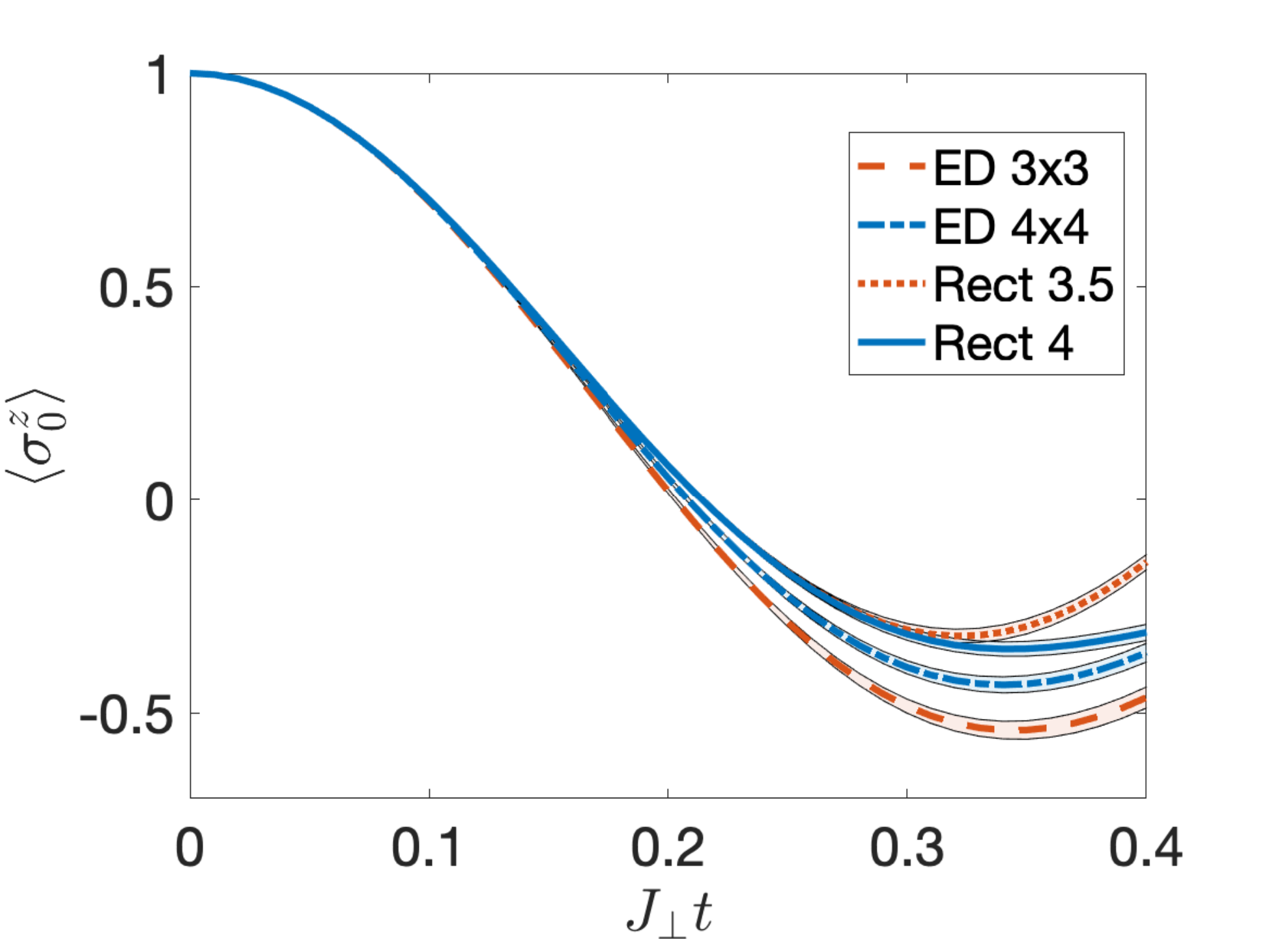}
		\caption{\label{fig:disorderaverage} Disorder-averaged time evolution of $\Braket{\sigma_0^z}$ for the system in Fig.~\ref{fig:sigma}(a), computed over ten independent disorder realizations using NLCE of orders 3.5 (orange dotted) and 4 (blue solid), and ED with $3\times 3$ (orange dashed) and $4\times 4$ (blue dash-dotted) lattices. The bands depict a deviation of one standard error of the mean, as computed from the ten realizations.}
	\end{figure}

	Figure~\ref{fig:disorderaverage} shows that the improved convergence of NLCE is not limited to a single disorder realization and remains when computing disorder averages. The system is the same as in Fig.~\ref{fig:sigma}(a), but the results are averaged over ten independent disorder realizations. Disorder averages are presented with deviations of one standard error of the mean. As in Fig.~\ref{fig:sigma}(a), the NLCE curves appear to diverge from each other at a later time than the ED curves, suggesting better convergence.

	\begin{figure}[htb]
		\includegraphics[width=.48\textwidth]{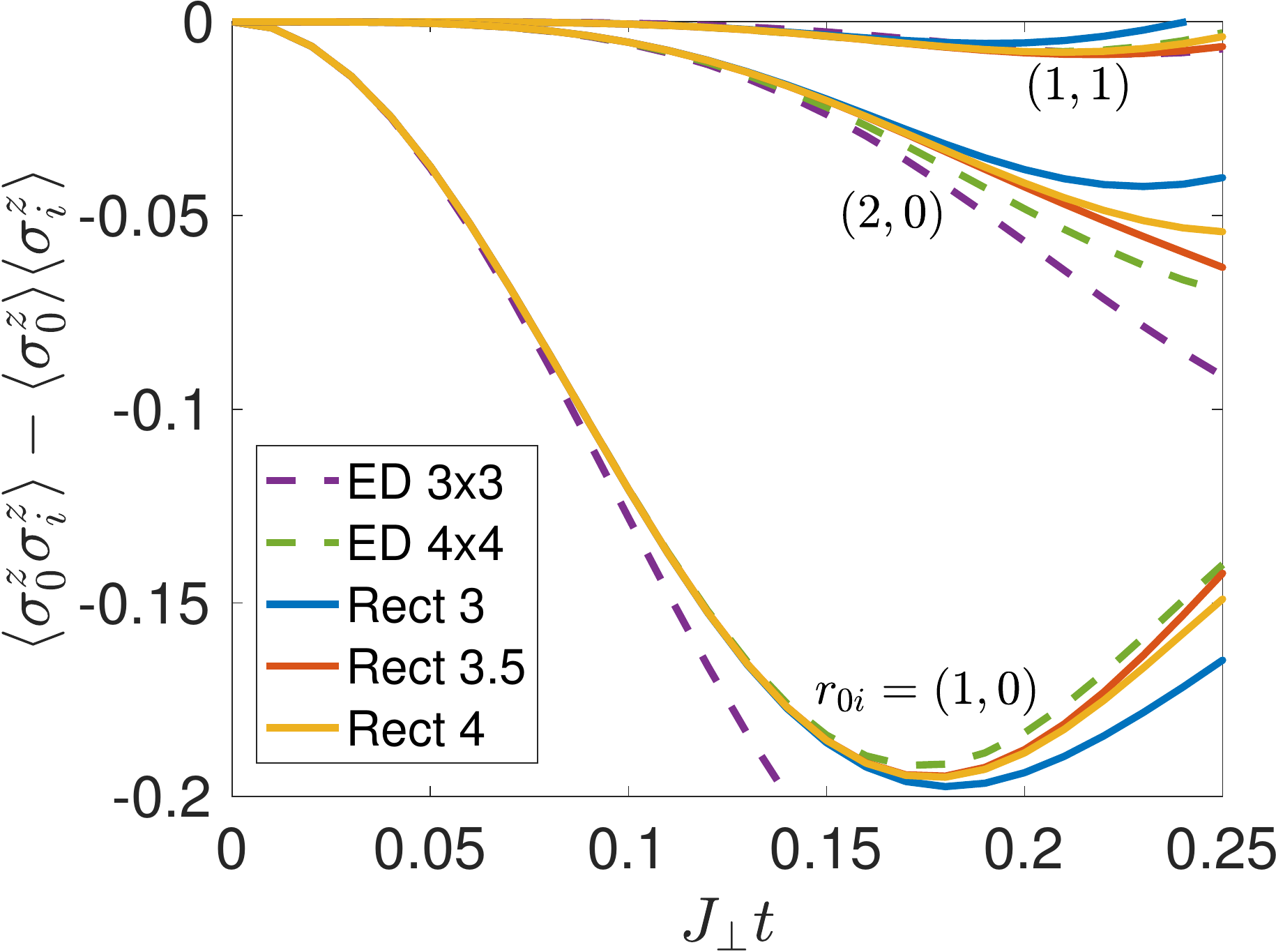}
		\caption{\label{fig:correlations} Two-point correlation functions of $\sigma^z$ at various separations $r_{0i} = (x_i, y_i)$ for the system in Fig.~\ref{fig:sigma}(a), computed for a single disorder realization. Solid curves are NLCE of orders 3 [blue: bottom for $(1, 0)$, top for rest], 3.5 [orange: top for $(1, 0)$, bottom for rest], and 4 (yellow: middle). Dashed curves are ED with $3\times 3$ [purple: top for $(1, 1)$ at short times, bottom for rest] and $4\times 4$ [green: bottom for $(1, 1)$ at short times, top for rest] lattices.}
	\end{figure}

	Figure~\ref{fig:correlations} shows that inhomogeneous NLCE also converges more rapidly than ED in computing observables such as correlation functions, which are spatially localized but not confined to a single site. The system is the same as in Fig.~\ref{fig:sigma}(a). Correlations $C_{0i}(t) = \Braket{\sigma_0^z\sigma_i^z} - \Braket{\sigma_0^z}\Braket{\sigma_i^z}$ between site 0 and site $i$ are computed for various separation vectors $r_{0i}$ between site 0 and site $i$. For all separations, NLCE is converged for longer times than ED at the same order, and---at least for separations of $(1, 0)$ and $(2, 0)$---even the order 3 NLCE seems to outperform the $4\times 4$ ED. It is interesting to note that the $(1, 1)$ correlation is lower in magnitude than the $(2, 0)$ correlation, despite the separation distance being smaller. This phenomenon appears only for a checkerboard initial state (not a uniform one), and can be explained by cancelations in perturbation theory that occur for the $(1, 1)$ correlation but not for the $(2, 0)$ correlation.

\begin{figure}[htb]
		\includegraphics[width=.48\textwidth]{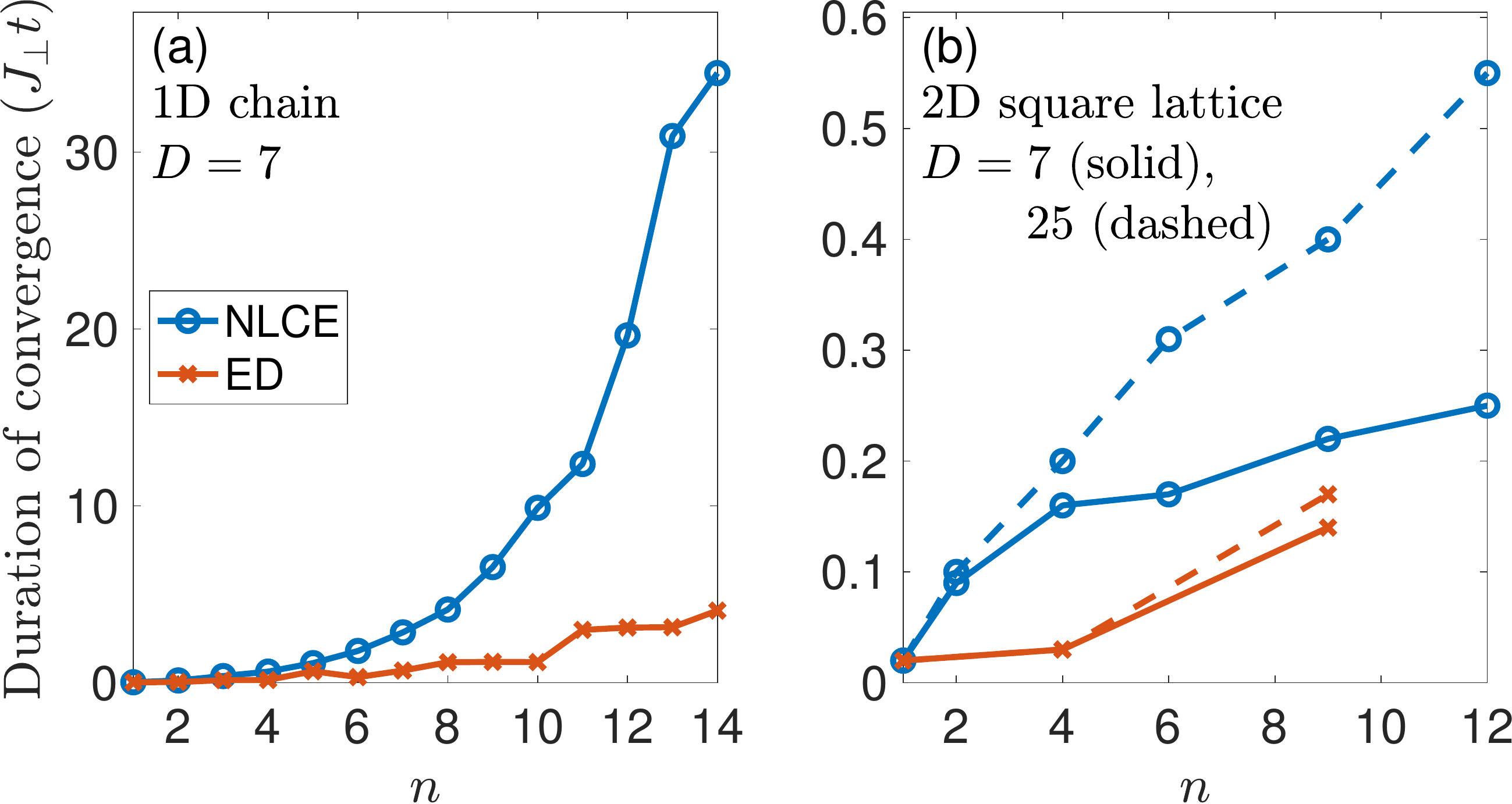}
		\caption{\label{fig:divergencetime} Duration of convergence as a function of simulation size for both NLCE and ED. System size $n$ is defined as in Fig.~\ref{fig:sigma}(d)-(f). Convergence for an NLCE of size $n$ is defined as earliest time at which the estimated $\Braket{\sigma_0^z(t)}$ deviates from the highest order NLCE estimate (order 15 for 1D, order 4 for 2D) by more than 1\% (results are qualitatively independent of this 1\% threshold). Convergence for an ED is defined similarly, but relative to the highest order ED estimate (15-site chain for 1D, $4\times 4$ lattice for 2D). The systems are (a) a 1D XXZ chain with disorder strength $D = 7J_\perp$, and (b) a 2D XXZ square lattice with disorder strength $D = 7J_\perp$ (solid) and $D = 25J_\perp$ (dashed).}
	\end{figure}

	The improved accuracy of NLCE over ED is greater for systems with more localized correlations, as in many-body localized systems. Figure~\ref{fig:divergencetime} shows the approximate duration of time for which NLCEs and EDs in one and two dimensions are well-converged. Figure~\ref{fig:divergencetime}(a) shows that in 1D, the duration of convergence of NLCE increases rapidly above $n \sim 10$, whereas ED shows a much slower convergence. A tentative interpretation of these results is that the convergence time grows rapidly with order once the clusters are large enough to capture the correlation length. The results are then consistent with the idea that NLCE can capture longer-ranged correlations than ED. Figure~\ref{fig:divergencetime}(b) shows that NLCE converges more slowly in 2D than in 1D, although it still converges rapidly relative to ED. We interpret this to mean that the 2D systems have localization lengths larger than we could capture in our simulations, or are not localized; this is consistent with the current understanding of many-body localization in dimensions above one. However, we do note that at higher disorder strength, NLCE appears to give a greater relative improvement in convergence over ED, as evidenced by the greater separation between the $D = 25$ curves than the $D = 7$ curves. This is consistent with higher disorder strength leading to a more localized system.

\section{\label{sec:conclusions}Conclusions}
	We have developed a generalized NLCE for simulating quantum systems without translational symmetry, including disordered systems and out-of-equilibrium systems with nonuniform initial conditions. We find that for dynamics computations with both Hamiltonian disorder and nonuniform initial conditions, our inhomogeneous NLCE provides more accurate results than an ED of comparable computation cost. This improvement can be seen for measurements in single disorder realizations, for disorder-averaged measurements, for measurements of single-site observables, and for measurements of multi-site observables such as two-point correlation functions. In our examples, increasing disorder strength increases the accuracy of NLCE relative to ED, which is consistent with the crucial role of the correlation length in determining the convergence of these numerical methods.

	In contrast to previous NLCE methods for disorder, one does not have to do a full (or nearly full) average over disorder configurations in order to obtain meaningful results. Whereas $m$-valued on-site disorder required summing $\mathcal O(m^N)$ (for the maximal number of sites $N$) disorder configurations to apply previous NLCE methods, ours requires a single disorder configuration---or a finite number to do disorder averaging with a finite statistical error---with an $\mathcal O(N)$ overhead. This method becomes especially useful for continuous disorder. Furthermore, cases that could not be treated with existing NLCE methods, for example the relaxation of an initially inhomogeneous configuration such as a domain wall, are possible with the new method.

	We have also provided a demonstration of the rectangle expansion NLCE in dynamics calculations. Our calculations show that the rectangle expansion gives intermediate accuracy for a given expansion order or computational cost: worse than the site expansion, but better than ED.

	The ideas presented here could potentially be used in tandem with other algorithms to improve its efficiency and accuracy. Firstly, resummation, which involves cleverly combining NLCE results across multiple orders to achieve a more accurate ``resummed'' result, has played a large role in the success of traditional NLCEs for computing low-temperature equilibrium properties~\cite{tang2013,rigol2006}. We have only presented so-called ``bare sum'' NLCE results that involve no resummation, and we speculate that an appropriate resummation scheme could greatly accelerate the order-by-order convergence of the inhomogeneous NLCE. However, preliminary results suggest that different resummation techniques are needed for dynamics than the popular Wynn and Euler resummations used in equilibrium. Secondly, nothing in the NLCE protocol explicitly requires the use of ED to compute property values $P(c)$. For example, combining NLCE with tensor network or quantum Monte Carlo methods for computing $P(c)$ could facilitate computing NLCEs of much higher order. This hybrid approach has already been applied in the homogeneous case with solvers like the density matrix renormalization group~\cite{kallin2014,stoudenmire2014,sahoo2016} and dynamical quantum typicality~\cite{richter2019}, and similar approaches are likely possible in the inhomogeneous case. In this sense, NLCE could be used as a general purpose ``convergence accelerator'' for any simulation algorithm applicable to arbitrary, finite-size clusters. This is especially true for the rectangle expansion; the site expansion involves summing a tremendous number of clusters at high orders, and is thus numerically unstable, but the rectangle expansion avoids this issue and is more robust to small errors from approximate solvers.

	After this paper was submitted, Ref.~\cite{richter2020} presented NLCE calculations for dynamics with a rectangle expansion. Although Ref.~\cite{richter2020} shares this feature with the present paper, it has a different focus. Our work focuses on inhomogeneity; while it uses a rectangle expansion, it relies on a basic cluster solver and does not study the expansion's particular features in depth. Ref.~\cite{richter2020} focuses more on the rectangle expansion, combines it with a more advanced cluster solver, and presents extensive comparisons to other state-of-the-art numerical methods.

\begin{acknowledgments}
	We would like to thank Ian White, Bhuvanesh Sundar, and Miroslav Hopjan for useful conversations. This material is based upon work supported with funds from the Welch Foundation Grant No. C-1872 and from NSF Grant No. PHY-1848304.
\end{acknowledgments}

\bibliography{nlcepaper-2020-05-06-jgan}

\end{document}